\documentclass[conference]{IEEEtran}[10pt]

 \makeatletter
\def\ifundefined{\@ifundefined}
\makeatother

\usepackage{times,amsmath}
\usepackage{graphicx}
\usepackage{amssymb}

\begin{document}
\thispagestyle{empty}

\title{\vspace{-0.2cm} \LARGE\bf Voice Service Support in Mobile Ad Hoc Networks\vspace{-0.3cm}
\thanks{This work was supported by the U.S. National Science Foundation under Grants
ANI-03-38807 and CNS-06-25637, and a Postdoctoral Fellowship and a
research grant from the Natural Science and Engineering Research
Council (NSERC) of Canada.}}

\author{
Hai Jiang$^\dag$, Ping Wang$^\S$, H. Vincent Poor$^\ddag$, and Weihua Zhuang$^\S$\\
$^\dag$Dept. of Elec. \& Comp. Eng., University of Alberta, Canada,
hai.jiang@ece.ualberta.ca\\
$^\S$Dept. of Elec. \& Comp. Eng., University of Waterloo, Canada,
\{p5wang, wzhuang\}@bbcr.uwaterloo.ca\\
$^\ddag$Dept. of Elec. Eng., Princeton University, USA, poor@princeton.edu\\
 \vspace{-10mm} }

   \providecommand{\IEEElabelindentfactori}{1.0}
   \providecommand{\IEEElabelindentfactorii}{0.75}
   \providecommand{\IEEElabelindentfactoriii}{0.0}
   \providecommand{\IEEElabelindentfactoriv}{0.0}
   \providecommand{\IEEElabelindentfactorv}{0.0}
   \providecommand{\IEEElabelindentfactorvi}{0.0}
   \providecommand{\labelindentfactor}{1.0}

   \providecommand{\iedlistdecl}{\relax}
   \providecommand{\calcleftmargin}[1]{
                   \setlength{\leftmargin}{#1}
                   \addtolength{\leftmargin}{\labelwidth}
                   \addtolength{\leftmargin}{\labelsep}}
   \providecommand{\setlabelwidth}[1]{
                   \settowidth{\labelwidth}{#1}}
   \providecommand{\usemathlabelsep}{\relax}
   \providecommand{\iedlabeljustifyl}{\relax}
   \providecommand{\iedlabeljustifyc}{\relax}
   \providecommand{\iedlabeljustifyr}{\relax}

   \newif\ifnocalcleftmargin
   \nocalcleftmarginfalse

   \newif\ifnolabelindentfactor
   \nolabelindentfactorfalse

   \newif\ifcenterfigcaptions
   \centerfigcaptionsfalse

    \let\OLDitemize\itemize
   \let\OLDenumerate\enumerate
   \let\OLDdescription\description

   \renewcommand{\itemize}[1][\relax]{\OLDitemize}
   \renewcommand{\enumerate}[1][\relax]{\OLDenumerate}
   \renewcommand{\description}[1][\relax]{\OLDdescription}

   \providecommand{\pubid}[1]{\relax}
   \providecommand{\pubidadjcol}{\relax}
   \providecommand{\specialpapernotice}[1]{\relax}
   \providecommand{\overrideIEEEmargins}{\relax}

     \let\CMPARstart\PARstart

   \let\OLDappendix\appendix
   \renewcommand{\appendix}[1][\relax]{\OLDappendix}

   \newif\ifuseRomanappendices
   \useRomanappendicestrue

\maketitle

\thispagestyle{empty}

\begin{abstract}
Mobile ad hoc networks are expected to support voice traffic. The
requirement for small delay and jitter of voice traffic poses a
significant challenge for medium access control (MAC) in such
networks. User mobility makes it more complex due to the
associated dynamic path attenuation. In this paper, a MAC scheme
for mobile ad hoc networks supporting voice traffic is proposed.
With the aid of a low-power probe prior to DATA transmissions,
resource reservation is achieved in a distributed manner, thus
leading to small delay and jitter. The proposed scheme can
automatically adapt to dynamic path attenuation in a mobile environment. Simulation
results demonstrate the effectiveness of the proposed scheme.
\end{abstract}

{\bf \it Keywords} -- \textbf{\small medium access control,
code-division multiple access, mobile ad hoc networks.}

\section{Introduction}\label{s:int}

Mobile ad hoc networks are expected to support multimedia services such as voice and video with quality-of-service (QoS) requirements. This task is challenging because there is no network central controller in such networks. User mobility makes the case even worse because of the time-varying network topology and propagation loss between any two nodes. QoS support in mobile ad hoc networks includes two directions: via routing at the network layer or via medium access control
(MAC) scheme at the link layer. The function of routing is
to search for a network path (i.e., a route) from a traffic source node to its destination node, and to react (e.g., select a new route) to possible route failures and/or network congestion. On the other hand, the function of MAC
is to coordinate the channel access in an orderly manner among the
mobile nodes such that efficiency can be achieved. Here, we focus primarily on the MAC in a mobile ad hoc network supporting
voice traffic. In such a network, a MAC scheme should be: 1) distributed (because of the scalability requirement); 2) tolerant to the hidden terminal problem; 3) ensuring of small delay/jitter (due to the real-time nature of voice traffic); and 4) adaptive to user mobility.
In this paper, we propose an effective MAC scheme with the above
desired features. The MAC scheme is fully distributed, thus scaling well
to a large network. The hidden terminal problem does not exist in
our scheme. Radio resource requirements are met by resource
reservation, thus achieving small delay and jitter. Our scheme can
also adapt to user mobility to a certain extent.

\section{Related Work and Discussion}\label{s:related_work}

The most popular MAC schemes for ad hoc
networks are carrier-sense multiple access (CSMA) and its
variants, and busy-tone based schemes. The CSMA schemes are
inherently distributed. However, they
suffer from the hidden terminal problem. 
Also, the schemes are based on an ideal assumption, i.e.,
that there is an interference range for each transmitter. As long as a
target receiver is outside the interference range of an
interferer, it is assumed that the interferer does not generate
any interference to the target receiver. However, in reality, the aggregate interference from a number of far-away interferers (each with a distance to the target receiver larger than the interference range) may corrupt the reception at the target receiver.
On the other hand, one motivation of busy-tone based schemes
\cite{Haas02} is to better solve the hidden terminal problem. With
the aid of a {\it receive busy-tone} transmitted by a target
receiver, the hidden terminals near the target receiver can be
notified not to transmit. However, the hidden terminal problem is
not eliminated. For instance, collisions of request-to-send (RTS)
frames can still happen due to the hidden terminal problem
\cite{Wang06}. It is also challenging to determine the busy-tone
coverage. Traditionally the busy-tone coverage is set to be the
same as the interference range. However, similar to the case in
CSMA, the aggregate
interference from multiple interferers outside the coverage of a busy-tone (sent by
a target receiver) can corrupt the reception at the target
receiver. In a mobile environment, the multipath fading can also
cause the busy-tone coverage to vary with time. This can severely
degrade the system performance because a strong interferer may not
be notified by the busy-tone due to instantaneous deep fading of
the busy-tone channel.

The contention-based nature of CSMA schemes and busy-tone based
schemes makes it very difficult to bound the delay and jitter
to small values for voice traffic. This observation
implies that we should instead resort to reservation of channel
resources. Resource reservation can avoid contentions,
and voice traffic can be transmitted continuously on the reserved
channel, thus making the transmission delay and jitter bounded.
Here we consider a code-division multiple access (CDMA)-based
mobile ad hoc network, where the air interface supports multiple
simultaneous transmissions in a neighborhood. In our previous work
\cite{Jiang_CCNC07}, we have proposed a MAC scheme for a
CDMA-based wireless mesh backbone, referred to as {\it MESH}
scheme here. Before DATA transmissions, each potential sender
first transmits a low-power probe to the network. Upon reception
of the probe, each active receiver estimates a potential
interference increase due to the potential sender's DATA
transmission. The active receiver transmits a busy-tone signal in
a separate busy-tone channel to notify the potential sender if the
potential interference increase is intolerable. If the potential sender does not detect a busy-tone, it
successfully reserves a code channel. However, targeted at a
wireless mesh backbone with fixed topology, the MESH scheme is not effective or efficient in a
mobile ad hoc network for the following reasons: 1) The power allocation strategy in MESH largely depends on the fixed network topology, and thus it is not applicable to a mobile network; 2) A single-frequency busy-tone channel is used in MESH. The busy-tone transmission can be affected significantly by fading and/or shadowing in a mobile environment. The coverage of the
busy-tone fluctuates with time, thus severely degrading the system
performance.

Based on the similar approach of ``probing" in the MESH
scheme, in the following we propose a MAC scheme for a mobile ad
hoc network supporting voice traffic.

\section{The Proposed MAC Scheme}\label{s:proposed}

Consider a mobile ad hoc
network with a number, $N$, of mobile nodes. At the MAC layer, a
source node can communicate with one or more of its neighbors,
with a separate queue for each destination node. Each node is
assigned a unique transmitting code and a unique receiving code
\cite{Sousa88}. Constant rate voice traffic is supported. The
network topology changes as the mobile nodes move.

Each voice link requires a minimum transmission rate (e.g., voice
packet generation rate from the codec) so that its delay and
jitter can be kept at a very low level. To keep the transmission
accuracy of the voice link, a signal to interference
plus noise ratio (SINR) threshold is required for each link,
i.e., for the link from sender $i$ to receiver $j$, the following
inequality should hold:
$\frac{G_iP_{ij}^tg_{ij}}{I_j+\eta_j}\ge \Gamma_{d}$,
where $G_i$ is the spreading gain, $P_{ij}^t$ is the transmit
power, $g_{ij}$ is the path gain from node $i$ to node $j$, $I_j$
is the interference (received at node $j$) from other links,
$\eta_j$ is the background noise power at node $j$, and
$\Gamma_{d}$ is the required SINR threshold for accurate information
transmission.

Unlike the situation for the MESH scheme, all the transmissions
are within a frequency band, using CDMA technology. A RAKE
receiver can collect signal energy from different paths. Hence, we
assume that there is no fast (multipath) fading, and the transmit
power is attenuated only due to path loss with exponent $\beta$
and shadowing. A node cannot transmit and receive at the same
time, since each node's transmission and reception are within the
same frequency band.

\begin{figure}
\begin{center}
\includegraphics[angle=0,width=0.5\textwidth]{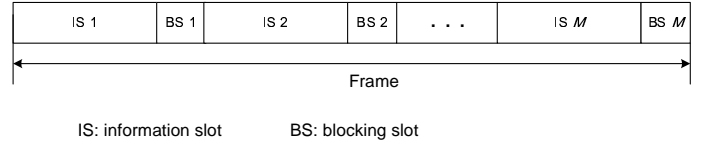}
\end{center}
\caption {Time frame structure.} \label{f:time_frame}
\end{figure}

In the time domain, we use a time frame structure as shown in Fig.
\ref{f:time_frame}. Time is partitioned into fixed-length frames.
In each frame, there are $M$ information slots (ISs), each
followed by a short blocking slot (BS). The ISs are used for
information (i.e., DATA and ACK) and probe transmissions, while
the BSs are used for the transmission of blocking messages (to be
discussed).

In a mobile ad hoc network, it is
difficult to estimate the path attenuation in advance from a
sender to a receiver due to user mobility. Therefore, the popular linear power allocation \cite{Moscibroda06} is not adopted. Rather, we consider
a common power allocation strategy, where each traffic source node
uses a constant power $P$ to transmit its DATA frames to its
destination, and the traffic destination uses the same power level
to feed back ACKs.

\subsection{Multiple Access Procedure}\label{s:mul_acc_pro}

\begin{figure}
\begin{center}
\includegraphics[angle=0,width=0.5\textwidth]{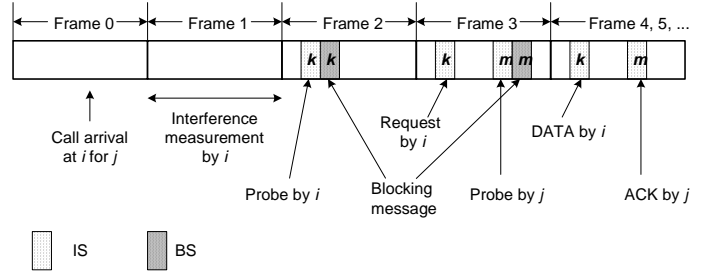}
\end{center}
\caption {The multiple access procedure.} \label{f:procedure}
\end{figure}

The multiple access capability of the network is based on resource
reservation. Consider the case of a new call arrival at source
node $i$ at time frame 0 to its destination node $j$. The multiple
access procedure is illustrated in Fig. \ref{f:procedure}.

\subsubsection{Selection of DATA IS}

Node $i$ first selects an IS for DATA transmission, referred
to as its {\it DATA IS}. The ISs at which the destination node $j$
is sending traffic should be skipped, because node $j$ cannot send
and receive at the same time at the same band. At frame 1, node
$i$ measures its experienced interference, and monitors the
transmitting code of node $j$. Among the ISs over which node $j$'s
transmitting code is not detected, node $i$ selects one with
the minimum interference level, say IS $k$, as its DATA IS.

\subsubsection{Transmissions of Probe and Blocking Message}

Prior to DATA transmission at the selected DATA IS, node $i$
should make sure its DATA transmission does not corrupt any
existing reception at any active receiver\footnote{Here an active
receiver can be an existing traffic destination node for DATA
reception, or an existing traffic source node for ACK reception.}
at the IS. To achieve this goal, node $i$ first sends a low-power
test signal into the network. Specifically, at IS $k$ of frame 2,
node $i$ sends a probe spread by a common probe code with a large
spreading gain, and with transmit power level $\alpha \cdot P$
($\alpha \ll 1$). No bit information is carried in the probe. With
the low power, the probe is very likely not to corrupt any
existing reception at the IS. Although the probe is sent with low
power, we assume that it can be received by active receivers at
the IS because of the large spreading gain.

At IS $k$, an active receiver, say node $l$, first measures the
received probe power level denoted by $P^{r,p}_l$ and its own
interference level due to other existing transmissions. It
estimates the potential interference increase due to DATA
transmission of the probe sender as $P^{r,p}_l/\alpha$, and
determines whether the potential interference increase can corrupt
its own reception. If so, it sends a blocking message at BS $k$.
The blocking message is spread by a common blocking code. No bit
information is carried by the blocking message. The power
allocation for the blocking message will be discussed in Section
\ref{s:bl_mes_pow}.

At BS $k$, node $i$ monitors the blocking code. If the detected
blocking message power is above a threshold $P^{r,b}_{th}$, it
gives up IS $k$ for DATA transmission and tries another IS until
no blocking message is detected.

\subsubsection{Selection of ACK IS}
If no blocking message is detected, node $i$ sends a request message to its destination node $j$, at
IS $k$ in frame 3 spread by node $j$'s receiving code, and sends
its first DATA frame at slot $k$ in frame 4 spread by node $i$'s
transmitting code, both with power $P$.

Upon receiving the request message, node $j$ selects an IS
for its ACK, referred to as {\it ACK IS}. Generally it is desired
that, at the ACK IS, the interference experienced by node $i$ is
small so that the ACK can be received correctly. So in the request
message, node $i$ indicates (among the ISs except IS $k$) the IS
with minimal experienced interference, say IS $m$. After node $j$
receives correctly the request, it sends an probe at IS $m$ in
frame 3\footnote{The probe is sent at IS $m$ in frame 3 when
$m>k$, or in frame 4 when $m<k$. For simplicity of presentation,
we use the former case as an example in this paper.} in the same
manner as node $i$ sending a probe for DATA IS. If no blocking
message is detected at BS $m$, node $j$ sends an ACK at IS $m$ in
frame 4 for the DATA frame received\footnote{Since the request
message and DATA are sent in the same way from node $i$, DATA can
also be received correctly if the request can be delivered
successfully.} in frame 4.

If node $i$ does not receive an ACK at IS $m$ of frame 4, it
selects another DATA IS, and repeats the above procedure with a
maximum number of attempts. If an ACK is received, it proceeds to
next step.

\subsubsection{Continuous Transmissions at Reserved DATA and ACK ISs}

At subsequent time frames from frame 5, node $i$ transmits its
DATA frames spread by its transmitting code at IS $k$, and node
$j$ transmits the ACKs spread by its transmitting code at IS $m$,
until the completion of the call. It can be seen that call-level
resource reservation is performed. Upon the call completion, no
link termination procedure is necessary, because our scheme is
based on real-time interference measurement, and other active
receivers can measure the interference
level reduction when the target link stops using the reserved resource.

\subsection{Power Allocation for the Blocking Message}\label{s:bl_mes_pow}
If an active receiver at an IS is to send a blocking message in
response to a probe from a potential sender\footnote{Here a
potential sender can be a traffic source node for DATA
transmission or a traffic destination node for ACK transmission.},
the transmit power of the blocking message should be designed
carefully so that the potential sender can be notified. Unlike the
MESH scheme, other active receivers at the IS do not need to hear
the blocking message. As the blocking message is sent in the same
frequency band as DATA frames, here we use a different method
(from that in the MESH scheme) for an active receiver to determine
its blocking message power based on the measured probe power. The
objective is to ensure that the potential sender can detect a
blocking message with power above threshold $P^{r,b}_{th}$.

At an IS, it is possible that one or more potential senders may
send probes for resource reservation. Since the resource
reservation is performed at the call level, it is reasonable to
assume that at an IS there are at most two potential probe senders
in a neighborhood. We also assume channel reciprocity in terms of
shadowing effect. Consider the case in which an active receiver,
say node $l$, detects the probe power level $P^{r,p}_l$, and wants
to send a blocking message. Node $l$ sends a blocking message with
transmit power
\begin{equation}\label{blk_mes_pwr}
P^{t,b}_{l} = \max\Big\{ \frac{2\cdot P^{r,b}_{th}\cdot \alpha
P}{P^{r,p}_l}, \frac{P^{r,b}_{th}\cdot P}{IM_l}\Big\}
\end{equation}
where $IM_l$ is the {\it interference
margin} \cite{Muqattash_dualch03} 
of node $l$, defined as the maximum extra interference that can be
tolerated by node $l$.

We first consider the scenario in which only one potential sender,
node $i$, sends a probe at an IS. The path gain from node $i$ to
node $l$ is
$g_{il} = \frac{P^{r,p}_l}{\alpha P}.$
From (\ref{blk_mes_pwr}) we have
\begin{equation}
P^{t,b}_{l} \ge \frac{2\cdot P^{r,b}_{th}\cdot \alpha
P}{P^{r,p}_l} =\frac{2\cdot P^{r,b}_{th}}{g_{il}}.
\end{equation}
Due to the channel reciprocity, $g_{li}=g_{il}$, we have
$P^{t,b}_{l} \ge \frac{2\cdot P^{r,b}_{th}}{g_{li}}.$
Then the received blocking message power at the potential sender
$i$ is
$P^{t,b}_{l}\cdot g_{li} \ge 2\cdot P^{r,b}_{th}$,
which is greater than the detection threshold $P^{r,b}_{th}$. So
the potential sender is notified to give up the IS.

Next, we consider the scenario in which two potential senders, node
$i_1$ and node $i_2$, send probes at an IS. As a common probe code is used, and no bit information is carried by a probe, node $l$ can collect energy from both probes via an RAKE receiver. The total received
probe power at node $l$ is also denoted by $P^{r,p}_l$. Suppose node
$l$ wants to send a blocking message. Without loss of generality,
assume the path gain from node $i_1$ to $l$ is larger than that
from $i_2$ to $l$, i.e., $g_{i_1l}\ge g_{i_2l}$. We have
\begin{equation}
P^{r,p}_l = \alpha P\cdot g_{i_1l}+\alpha P\cdot g_{i_2l} \le
2\alpha P\cdot g_{i_1l}.
\end{equation}
When node $l$ sends a blocking message with power given by
(\ref{blk_mes_pwr}), the received blocking message power at node
$i_1$ is
\begin{equation}
\begin{array}{ll}
P^{t,b}_{l}\cdot g_{li_1} &= P^{t,b}_{l}\cdot g_{i_1l}\\
&\ge \frac{2\cdot P^{r,b}_{th}\cdot \alpha P}{P^{r,p}_l} \cdot g_{i_1l}\\
& \ge \frac{2\cdot P^{r,b}_{th}\cdot \alpha P}{2\alpha P\cdot g_{i_1l}} \cdot g_{i_1l}\\
&=P^{r,b}_{th}.
\end{array}
\end{equation}
So node $i_1$ will be notified to give up the IS.

In the following, we prove that, if node $i_2$'s potential
information (i.e., DATA or ACK) transmission generates extra
interference that is intolerable at node $l$, node $l$'s blocking
message with power given by (\ref{blk_mes_pwr}) can reach node
$i_2$ with a power level above the threshold $P^{r,b}_{th}$.

\begin{proof}
If node $i_2$'s information transmission can corrupt node $l$'s
reception, this means
$P\cdot g_{i_2l} \ge IM_l.$
We have
$g_{i_2l} \ge \frac{IM_l}{P}.$
When node $l$ sends a blocking message with power given by
(\ref{blk_mes_pwr}), the received blocking message power at node
$i_2$ is
\begin{equation}
\begin{array}{ll}
P^{t,b}_{l}\cdot g_{li_2} &= P^{t,b}_{l}\cdot g_{i_2l}\\
&\ge \frac{P^{r,b}_{th}\cdot P}{IM_l} \cdot g_{i_2l}\\
&\ge \frac{P^{r,b}_{th}\cdot P}{IM_l} \cdot \frac{IM_l}{P}\\
&=P^{r,b}_{th}.
\end{array}
\end{equation}
As node $i_2$ detects blocking message power above the threshold,
it gives up the IS.
\end{proof}

\subsection{Impact of User Mobility and Solutions}
The impact of user mobility on each active link at the MAC layer
is two-fold. One is due to the time-varying path attenuation of
the desired signal, and the other is due to the varying
interference level. The path attenuation consists of three
components: path loss, shadowing, and fast (multipath) fading.
Fast fading can be addressed by the RAKE receiver that collects
signal energy from different paths. The time-varying path loss,
shadowing, and interference level make the SINR of each link
fluctuate with time. Generally our scheme can automatically adapt
to the varying SINR of each link, because the receiver makes
real-time measurement of its desired signal and interference
levels. However, it is possible that the SINR of some ongoing
links at an IS cannot remain above the threshold $\Gamma_d$ (a
condition referred to as a {\it link failure}) because of the
fluctuations. Although a new route searched by the routing scheme
can help to solve the problem, it is desirable to deal with the
problem at the MAC layer first before resorting to the routing
scheme, considering the overhead and time needed for a new route
search, and the duration needed to set up a new route.

To address the problem, for each active link, the traffic
destination node indicates in its ACK two candidate DATA ISs at
which the destination node experiences the minimum interference
and does not transmit. The destination node also indicates a detected link
failure (if any) in the ACK. If the traffic source node (say node $i$) does
not receive correctly the ACK from its destination (say node $j$),
this means a link failure happens because the ACK is
corrupted\footnote{If an expected DATA frame is not received, the
traffic destination node responds with a NACK. For presentation
simplicity, we use ``ACK" to represent both ACK and NACK in this
paper.}. As node $i$ does not have the information of whether the
DATA transmission is corrupted or not, we let the link be
re-established at a new DATA IS and a new ACK IS. The main
difference from the establishment procedure given in Section
\ref{s:mul_acc_pro} is as follows. Node $i$ sends two probes at
the two candidate DATA ISs indicated in the previous ACK. At each
candidate IS, a request is sent if no blocking message is
detected. The request also indicates two candidate ACK ISs at
which node $i$ experiences the minimum interference and does not transmit. If at least
one request is received at node $j$ (which means a new DATA IS is
established), node $j$ sends two probes at the candidate ACK ISs.
If at least one probe at a candidate ACK IS is not blocked by
active receivers at that IS, a new ACK IS is established. If either
a new DATA IS or a new ACK IS cannot be established, the call is
dropped.

On the other hand, if node $i$ successfully receives an ACK which
indicates that a failure of DATA transmission happens, a similar
procedure is executed, except that only the establishment of a new
DATA IS is needed.

Generally, if the link failure is due to large interference, it is
likely that the call can be established successfully in other DATA
and/or ACK ISs at which nodes generate large interference are not active.
However, if the link failure is due to a large path loss of the
desired signal (e.g., the traffic source and destination are
separated by a large distance) or due to shadowing of the desired
signal (e.g., the path between the traffic source and destination
is blocked by a tall building), it is very likely that the link
quality of the target call at other ISs is not good enough either, and
thus the call will be dropped. In this case, a new route should be
selected by the routing scheme.


The proposed MAC scheme has the desired features discussed in
Section \ref{s:int}. The scheme is performed in a distributed
manner at the mobile nodes, thus scalable to large networks. The
hidden terminal problem does not exist, since each
active receiver makes an admission decision on a new call based on
real-time measurement of its own desired signal and interference
levels. The reservation nature of our scheme can guarantee small
delay and jitter for voice traffic as long as the minimum required
rate can be met. Our scheme can adapt automatically to user
mobility to a certain extent, also benefiting from the fact that
each active receiver makes real-time measurement of desired signal
and interference levels. In addition, the ``ideal assumption" required by CSMA and busy-tone based schemes (as discussed in Section \ref{s:related_work}) is unnecessary for our scheme. Furthermore,
although the blocking message has similar functionality to the
busy-tone in the busy-tone based scheme, our scheme does not have the
problem of different channel gains in separate busy-tone frequency band and
DATA frequency band, because the blocking message and DATA are
sent in the same band.

\section{Performance Evaluation}\label{s:performance}
We run computer simulations to evaluate the performance of the
proposed MAC scheme. Since MAC mainly deals with the transmission from a node to its neighbors, only one-hop transmission is considered here. Consider an ad hoc network consisting of $N$
nodes, with $N/2$ voice source nodes and $N/2$ voice destination
nodes. The source nodes are randomly distributed in a 1000 meter
$\times$ 1000 meter square, each associated with a destination
node. All the source nodes are fixed, and all the destination
nodes move with a velocity $v$ and a randomly selected direction.
In order to maintain a satisfactory SINR, a destination node is
assumed to move within a radius of 150 m of the associated source
node\footnote{Routing may be involved when a destination node moves beyond this distance, a situation which is beyond the scope of this work.}. Each time frame has a fixed duration of 20
ms, and is further divided into 8 ISs and 8 BSs. The spreading
gain is 32 for DATA and ACKs, and 3200 for the probe. The path
loss attenuation exponent $\beta$ is 2.4, and the shadowing is
modeled by the first-order autoregressive process given in \cite{Stuber96}. Voice call inter-arrival time at each
source node is exponentially distributed with mean value 48 seconds,
and the duration of a call is exponentially distributed with mean
value 30 seconds. When a call is active, one DATA frame is
generated over each time frame. The SINR requirement $\Gamma_d$ is 10 dB.

In the simulations, the number $N$ of nodes varies from 20 to 100,
and the destination node velocity $v$ varies from 0 to 20 km/hour (kph). We
first verify our assumption in Section \ref{s:bl_mes_pow} that the
probability of more than two probe senders at an IS is negligible.
Among the ISs with one or more probe senders, Table
\ref{tab.probe_prob} lists the percentage of ISs that have one
probe sender, two probe senders, and more than two
probe senders, respectively. No more than two
probes are observed at an IS, thus validating our assumption.

\begin{table}
\begin{center}
\caption{The percentage of the ISs with different number of
probes.}\label{tab.probe_prob}
\begin{tabular}{c|c|c|c|c|c|c}
\hline\hline \multicolumn{2}{c|}{Number ($N$) of nodes} & 20 &40 &60& 80& 100\\
\hline
 &   1 probe & 1.000  & 1.000 &   0.998 &   0.996  &  0.997\\
\cline{2-7} \raisebox{1.5ex}[0cm][0cm]{$v=10$}&   2 probes & 0 & 0 & 0.002 & 0.004 & 0.003\\
\cline{2-7} \raisebox{1.5ex}[0cm][0cm]{kph}&   $\ge 3$ probes & 0 & 0 & 0& 0& 0\\
\hline
 &   1 probe & 1.000  &  0.996  &  1.000  &  0.995  &  0.996\\
\cline{2-7} \raisebox{1.5ex}[0cm][0cm]{$v=20$} &   2 probes & 0  & 0.004 & 0 & 0.005 & 0.004 \\
\cline{2-7} \raisebox{1.5ex}[0cm][0cm]{kph} &   $\ge 3$ probes & 0 & 0 & 0& 0& 0\\
\hline\hline
\end{tabular}
\end{center}
\end{table}

\begin{figure}
\begin{center}
\includegraphics[angle=0,width=0.5\textwidth]{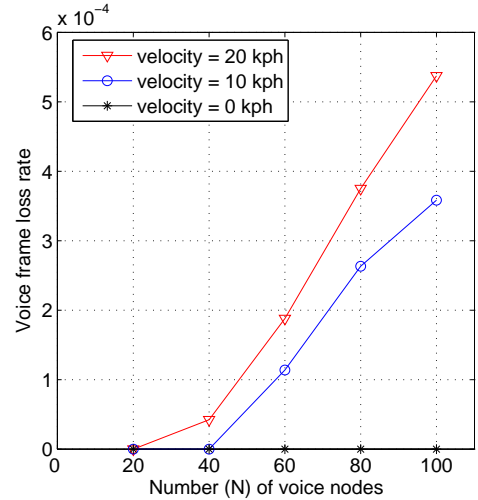}
\end{center}
\caption {The DATA frame loss rate of a voice call.} \label{f:frame_drop}
\end{figure}

For an ongoing voice call, if its SINR requirement is not
satisfied, it tries to switch from one IS to another. Some DATA
frames will be lost during the switching due to the switching
processing time, and if the switching is not successful, call
dropping occurs. Fig. \ref{f:frame_drop} shows the DATA frame loss rate
with different $N$ and $v$ values. A larger value of $N$ and/or $v$
tends to increase the frequency of switching among the ISs, thus
increasing the frame loss rate. It can also be seen that all the
frame loss rates are bounded by 0.05\%, which is well within acceptable range for voice. It is also observed that the call dropping
rate is bounded by 1.2\%, which happens when $N=100$ and $v=20$
kph. So our scheme can keep the frame loss rate and call dropping
rate at a low level.

\section{Conclusions and Discussions}\label{s:con}
It is challenging to support voice traffic over a mobile ad hoc
network due to the delay-sensitive nature of such traffic and due
to user mobility. In this paper we have proposed a MAC scheme to
support voice traffic
by the resource reservation mechanism. User mobility is
addressed by real-time signal and interference measurements at the
receivers. This study should provide helpful insight to the
development and deployment of mobile ad hoc networks supporting
multimedia services.

Currently constant rate voice traffic has been considered in this
research. However, voice traffic is characterized by an {\tt
on-off} nature, and no information packets are generated at an {\tt
off} state. When an {\tt
on} voice flow becomes {\tt off}, other active receivers measure
less interference. If subsequently one or more new calls are
admitted at the serving ISs of the {\tt off} voice flow, and later
the {\tt off} voice flow switches to the {\tt on} state, it is likely
that some active receivers experience more interference than what they can tolerate, thus leading to link corruption. To address this problem, when a
voice flow becomes {\tt off}, its source and destination nodes
continuously send (on the respective serving IS) a probe (with a
different code from that used in Section \ref{s:mul_acc_pro}) with a
low power and a large spreading gain. Through the received probe
power, an active receiver at the IS can estimate the potential
interference level which will be generated if the {\tt off} voice
flow becomes {\tt on} again. The potential interference is counted
in the active receiver's current interference level. We call this method {\it 100-percent reservation}, without a multiplexing gain. Research on statistical multiplexing of voice traffic within this scheme is currently underway.


\begin{thebibliography}{99}

\bibitem{Haas02}
Z. J. Haas and J. Deng, ``Dual busy tone multiple access (DBTMA)
-- a multiple access control scheme for ad hoc networks," {\it
IEEE Trans. Commun.}, vol. 50, no. 6, pp. 975--985, June 2002.

\bibitem{Wang06}
P. Wang, H. Jiang, and W. Zhuang, ``A dual busy-tone MAC scheme
supporting voice/data traffic in wireless ad hoc networks," in
{\it Proc. IEEE GLOBECOM'06}, San Francisco,
California, USA, Nov.--Dec. 2006.

\bibitem{Jiang_CCNC07}
H. Jiang, P. Wang, W. Zhuang, and X. Shen, ``An interference aware
distributed MAC scheme for CDMA-based wireless mesh backbone," in
{\it Proc. IEEE CCNC'07}, pp. 59--63, Las Vegas, Nevada, USA, Jan.
2007.

\bibitem{Sousa88}
E. S. Sousa and J. A. Silvester, ``Spreading code protocols for
distributed spread-spectrum packet radio networks," {\em IEEE
Trans. Commun.}, vol. 36, no. 3, pp. 272--281, Mar. 1988.





\bibitem{Moscibroda06}
T. Moscibroda and R. Wattenhofer, ``The complexity of connectivity
in wireless networks," in {\it Proc. IEEE INFOCOM'06}, Barcelona, Spain, Apr. 2006.





\bibitem{Muqattash_dualch03}
A. Muqattash and M. Krunz, ``Power controlled dual channel (PCDC)
medium access protocol for wireless ad hoc networks," in {\it
Proc. IEEE INFOCOM'03}, pp. 470--480, San Francisco, California,
USA, Mar.--Apr. 2003.

\bibitem{Stuber96}
G. L. Stuber, {\it Principles of Mobile Communication}. Kluwer
Academic Publishers, Norwell, MA, USA, 1996, pp. 89-90.

\end{thebibliography}
\end{document}